# Damping Rate of a Scalar Particle in Hot Scalar QED[*]

Markus H. Thoma and Christoph T. Traxler

*Institut für Theoretische Physik, Universität Giessen,*

*35392 Giessen, Germany*

## Abstract

In contrast to the damping of partons in a quark-gluon plasma, the damping of a scalar particle in a hot scalar QED plasma can be calculated to leading order for the whole momentum range using the Braaten-Pisarski method. In this way the evolution of the logarithmic infrared singularity caused by the exchange of a transverse photon from soft to hard momenta can be studied.

The damping rates of quarks and gluons in a quark-gluon plasma (QGP) have been studied in a number of papers (see ref. [1] and references quoted there), because they are of theoretical as well as practical interest. For instance, the gluon damping rate at zero momentum, calculated in lowest order perturbation theory, turned out to be gauge dependent ("plasmon puzzle") [2], stimulating the development of an improved perturbation theory (Braaten-Pisarski method [3]). Furthermore, this rate is related to the Lyapunov exponent of the classical Yang-Mills theory [4]. The damping rates of hard partons with momenta of the order of the temperature or larger, on the other hand, provide a simple test for infrared problems in finite temperature gauge theories (see e.g. [5]) and are closely connected to interesting quantities of the QGP such as the thermalization time in ultrarelativistic heavy ion collisions [6], the energy loss of energetic partons [7,8] and transport coefficients [9].

A consistent calculation of damping rates in hot gauge theories requires the use of the Braaten-Pisarski method, leading to gauge invariant results complete to leading order in the coupling constant. This method is based on the resummation of a class of diagrams, called hard thermal loops (HTL), from which effective Green's functions are constructed. These effective Green's functions have to be used if all external legs of the Green's function under consideration are soft, i.e. of the order of $gT$, where the coupling constant $g$ is assumed to be small.

---

[*]Supported by BMBF and GSI Darmstadt

Damping rates, defined by the imaginary part of dispersion relations, are proportional to the imaginary part of self energies [10]. In the case of the damping rate of a soft gluon we have to consider the self energy shown in fig.1, where the blobs denote resummed Green's functions. Since the external legs of these diagrams are soft, effective propagators as well as vertices have to be adopted. Due to the complicated structure of these effective Green's functions, in particular of the effective vertices, the damping rate of a soft gluon has been calculated only at zero momentum, yielding the gauge invariant result [11]

$$\gamma_g(0) = 0.2640 \; g^2 \; T. \tag{1}$$

The physical mechanism of this damping can be seen from cutting the diagrams of fig.1 [12] resulting in graphs that correspond to elastic scattering and gluon bremsstrahlung.

In the case of a hard gluon the damping rate is determined from the imaginary part of the self energy in fig.2, where only one effective propagator and no effective vertices are needed, as the external lines of the diagram are hard. Cutting this diagram, we see that the damping is due to elastic scattering, but there is no gluon bremsstrahlung contribution to leading order. The hard damping rate exhibits a logarithmic infrared divergence, caused by the absence of static magnetic screening in the transverse part of the effective gluon propagator. At zero momentum the transverse propagator, corresponding to a chromomagnetic interaction, does not contribute and the Debye screening in the longitudinal (chromoelectric) part is sufficient to provide a finite result. Assuming an infrared cutoff from higher order contributions beyond the Braaten-Pisarski resummation scheme, for example by a non-perturbative magnetic mass of the order $g^2T$ [14] or the damping itself [15], the following result is obtained [13]

$$\gamma_g(p \gtrsim T) = \frac{3g^2 T}{4\pi} \ln \frac{c}{g}, \tag{2}$$

where the constant $c$ cannot be determined using the Braaten-Pisarski method and will be chosen as $c = 1$ in the following (logarithmic approximation).

The same observation holds for the damping rate of a quark: At zero momentum a finite result is found [16,17], whereas for hard momenta a logarithmic infrared divergence shows up [7,5].

It would be of interest to study the transition from zero momentum to hard momenta in the damping rates, i.e., how the infrared divergence switches on. Unfortunately, the calculation of the soft damping rates in QCD at non-zero momenta is very involved due to the complicated momentum dependence of the effective vertices in fig.1. In order to illuminate this problem, we consider scalar QED as a toy model for QCD [18]. We expect to gain insight into the development of the infrared singularity in the gluon damping rate with increasing momentum by studying the damping rate of a scalar particle in scalar QED as function of the momentum, because the structure of both theories at high temperatures is similar. For example, the effective gauge boson propagators show the same momentum dependence. However, there are no effective vertices in scalar QED and the effective scalar HTL self energy is momentum independent, facilitating the computation of the scalar damping rate for arbitrary momenta enormously. (The damping rate of a photon is infrared finite even for hard momenta [19] and thus cannot be used to study the above problem.)

As a first step of the calculation of the scalar damping rate we have to determine the effective Green's functions in scalar QED. The effective photon propagator in Coulomb gauge follows from the resummation of the HTL photon self energy of fig.3 [18],

$$\begin{aligned} D_L^{\star -1} &= k^2 + 3\, m_\gamma^2 \left(1 - \frac{k_0}{2k} \ln \frac{k_0 + k}{k_0 - k}\right), \\ D_T^{\star -1} &= k_0^2 - k^2 - \frac{3}{2} m_\gamma^2 \left[1 - \left(1 - \frac{k^2}{k_0^2}\right) \frac{k_0}{2k} \ln \frac{k_0 + k}{k_0 - k}\right], \end{aligned} \tag{3}$$

where $m_\gamma = eT/3$ and $k = |{\bf k}|$. In QCD the effective gluon propagator differs from the effective photon propagator (3) only by replacing $m_\gamma^2$ by $m_g^2 = g^2 T^2(1 + N_f/6)/3$, where $N_f$ is the number of thermalized quark flavors in the QGP.

The effective scalar propagator is obtained by resumming the HTL scalar self energy diagrams of fig.4 [18], resulting in the simple expression

$$\Delta^{\star -1} = k_0^2 - k^2 - m_s^2 \qquad (4)$$

with $m_s = eT/2$.

The damping rate of a charged scalar particle is given by

$$\gamma_s(p) = -\frac{1}{2\omega_p} Im\, \Xi^\star(p_0 = \omega_p, p), \qquad (5)$$

where $\omega_p = \sqrt{p^2 + m_s^2}$ and the scalar self energy $\Xi^\star$ is shown in fig.5. Cutting this diagram corresponds to elastic scattering of scalar particles off scalar particles via the exchange of a photon. The decay of a scalar particle into a scalar particle plus a photon is kinematically forbidden. As opposed to gluon damping there is no bremsstrahlung contribution for soft momenta.

It is convenient to decompose the scalar damping rate into a part $\gamma_L$ corresponding to the exchange of a longitudinal photon and a part $\gamma_T$ corresponding to the exchange of a transverse photon: $\gamma_s = \gamma_L + \gamma_T$. Using the effective propagators of (3) and (4) for calculating the diagram of fig.5 within the imaginary time formalism we find

$$\gamma_L(p) = \frac{e^2}{16\pi} \frac{1}{\omega_p} \int_0^\infty dq\, q^2 \int_{-1}^1 d\eta\, \frac{(\omega_p + \omega_k)^2}{\omega_k} \left[n(\omega_k) + n(\omega_p - \omega_k)\right] \rho_L(\omega_p - \omega_k, q),$$

$$\gamma_T(p) = \frac{e^2}{4\pi} \frac{p^2}{\omega_p} \int_0^\infty dq\, q^2 \int_{-1}^1 d\eta\, \frac{1-\eta^2}{\omega_k} \left[n(\omega_k) + n(\omega_p - \omega_k)\right] \rho_T(\omega_p - \omega_k, q), \qquad (6)$$

where $q$ is the magnitude of the three momentum of the exchanged photon, $k = |\mathbf{p} - \mathbf{q}|$ the one of the internal scalar particle and $\eta$ the cosine of the angle between $\mathbf{p}$ and $\mathbf{q}$. Bose distributions are given by $n(\omega) = 1/[\exp(\omega/T) - 1]$ and the spectral functions of the effective photon propagator by [20]

$$\rho_{L,T}(\omega, q) = -\frac{1}{\pi} Im\, D_{L,T}(\omega + i\epsilon, q). \qquad (7)$$

The integrals over $q$ and $\eta$ in (6) have to be performed numerically. The results for $\gamma_L(p)/e^2 T$ and $\gamma_T(p)/(e^2 T \ln 1/e)$ as functions of $p/T$ are shown in fig.6, where we have chosen $e = 0.01$ in order to guarantee the weak coupling limit. We note that the damping rate depends only weakly upon the momentum over the whole range except from momenta of the order $eT$.

Now we discuss the limiting cases $p \to 0$ and $p \gtrsim T$. In the first case, we may set $\omega_p = m_s$ and $k = q$. Furthermore, $q$ as well as $k$ are soft and we replace the distribution functions by $T/\omega_k$ and $T/(\omega_p - \omega_k)$. Then the $\eta$-integration becomes trivial. The remaining integration over $q$ has to be done numerically, yielding for the longitudinal part of the damping rate

$$\gamma_L(0) = 0.02870 \, e^2 \, T. \tag{8}$$

(The small deviation of (8) from the numerical solution of (6) in fig.6 at small momenta is caused by the use of the approximation $n(\omega) \simeq T/\omega$ for the distribution functions.)

The transverse part is logarithmically infrared divergent. Therefore the momentum integration is dominated by small $q$. For $q \ll eT$ we approximate $\omega = \omega_p - \omega_k \simeq q^2/(2m_s)$. Thus we may use the static limit, $|\omega| \ll q$, for the transverse spectral function [14],

$$\rho_T(\omega, q) = \frac{3m_\gamma^2 \omega q}{4} \frac{1}{q^6 + (3\pi m_\gamma^2 \omega/4)^2}. \tag{9}$$

Then the logarithmically divergent part can be extracted analytically:

$$\gamma_T(p \to 0) = \frac{16 e^2 T}{9\pi^3} \frac{p^2}{m_\gamma^2} \ln \frac{1}{e}, \tag{10}$$

where a non-perturbative infrared cutoff of the order $e^2 T$ has been assumed. As in the case of quarks and gluons the transverse contribution to the damping rate vanishes at zero momentum. So the total damping rate of a scalar particle at zero momentum is given by (8). For small momenta the logarithmically divergent part of the damping rate is switched on proportional to $p^2/m_\gamma^2$ and is of higher order compared to (8) as long as $p \lesssim e^2 T$.

For hard momenta $p \gtrsim T$, we may set $m_s = 0$, $k = p \gg q$, $\omega_p - \omega_k = q\eta$, $n(\omega_k) = 0$, and $n(\omega_p - \omega_k) = T/(\omega_p - \omega_k)$. Then the integration over $q$ can be done exactly, while the one over $\eta$ has to be performed numerically, yielding

$$\gamma_L(p \gtrsim T) = 0.04369 \, e^2 \, T, \tag{11}$$

which agrees, apart from color factors, with the result for quarks and gluons and is independent of the momentum $p$.

In the case of the transverse part we obtain analogously to the QCD case [5] the logarithmic divergent result

$$\gamma_T(p \gtrsim T) \simeq \frac{e^2 T}{4\pi} \ln\frac{1}{e}, \tag{12}$$

where we assumed an infrared cutoff of the order $e^2 T$ again. (In this calculation the static limit (9) has been assumed, which leads to a result differing by about 10% from the one obtained with the full spectral function (7), as can be seen by comparing with fig.6.)

In conclusion, we have calculated the damping rate of a scalar particle in scalar QED for the whole momentum range to leading order using the effective perturbation theory developed by Braaten and Pisarski. The logarithmic infrared singularity in the part coming from the exchange of a transverse photon, is switched on with a factor $p^2/m_\gamma$ for small $p \ll eT$. In order to avoid this divergence a treatment beyond the Braaten-Pisarski method is required.

An application of the calculation presented here is the damping rate of a Higgs particle in an electroweak plasma above the phase transition. This damping rate is almost identical to the one of a scalar particle in scalar QED and will be presented in a forthcoming publication [21].

## ACKNOWLEDGMENTS

We would like to thank A. Rebhan for helpful discussions.

# REFERENCES

[1] M.H. Thoma, to be published in: *Quark-Gluon Plasma II*, ed. R. Hwa (World Scientific), hep-ph/95

[2] J.A. Lopez, J.C. Parikh, and P.J. Siemens, Texas A&M preprint (1985) unpublished

[3] E. Braaten and R.D. Pisarski, Nucl. Phys. B337 (1990) 569

[4] T.S. Biró, C. Gong, B. Müller, and A. Trayanov, Int. Jour. Mod. Phys. C5 (1994) 113

[5] C.P. Burgess and A.L. Marini, Phys. Rev. D45 (1992) R17; A. Rebhan, Phys. Rev. D46 (1992) 482

[6] M.H. Thoma, Phys. Rev. D49 (1994) 451

[7] M.H. Thoma and M. Gyulassy, Nucl. Phys. B351 (1991) 491

[8] E. Braaten and M.H. Thoma, Phys. Rev. D44 (1991) R2625

[9] M.H. Thoma, Phys. Lett. B269 (1991) 144

[10] U. Heinz, K. Kajantie, and T. Toimela, Ann. Phys. (N.Y.) 176 (1987) 218

[11] E. Braaten and R.D. Pisarski, Phys. Rev. D42 (1990) 2156

[12] H.A. Weldon, Phys. Rev. D28 (1983) 2007

[13] E. Braaten, Nucl. Phys. B (Proc. Suppl.) 23B (1991) 351

[14] R.D. Pisarski, Phys. Rev. Lett. 63 (1989) 1129

[15] V.V. Lebedev and A.V. Smilga, Ann. Phys. (N.Y.) 202 (1990) 229

[16] E. Braaten and R.D. Pisarski, Phys. Rev. D46 (1992) 1829

[17] R. Kobes, G. Kunstatter, and K. Mak, Phys. Rev. D45 (1992) 4632

[18] U. Kraemmer, A.K. Rebhan, and H. Schulz, Ann. Phys. (N.Y.) 238 (1995) 286

[19] M.H. Thoma, Phys. Rev. D51 (1995) 862

[20] R.D. Pisarski, Physica A158 (1989) 146

[21] T.S. Biró and M.H. Thoma, in preparation

FIGURES

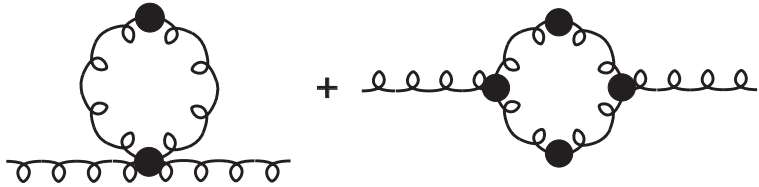

FIG. 1. Gluon self energy determining the damping rate of a soft gluon.

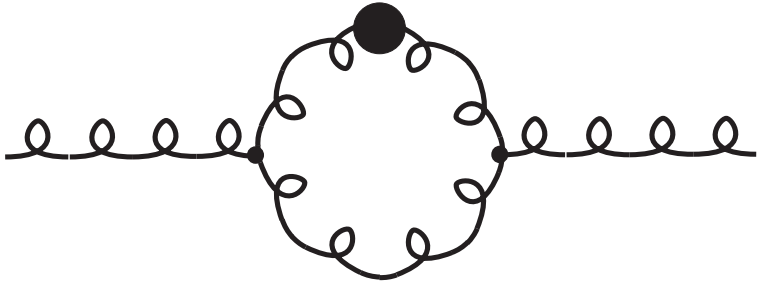

FIG. 2. Gluon self energy determining the damping rate of a hard gluon.

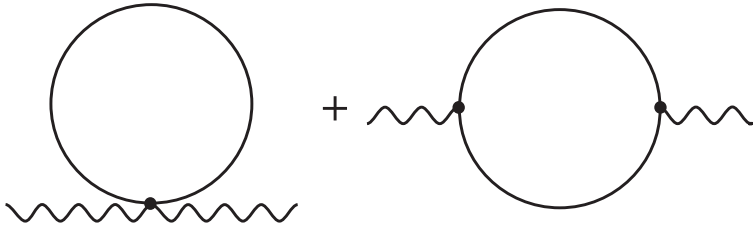

FIG. 3. Hard thermal loop diagrams of the photon self energy.

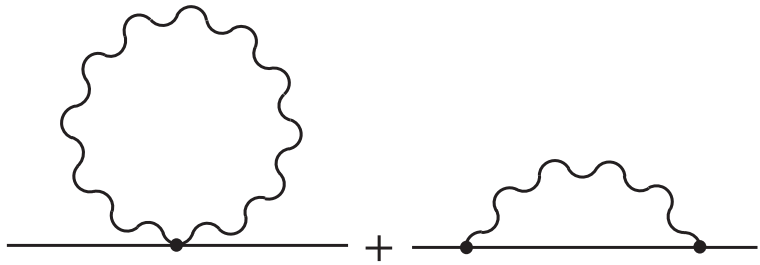

FIG. 4. Hard thermal loop diagrams of the scalar self energy.

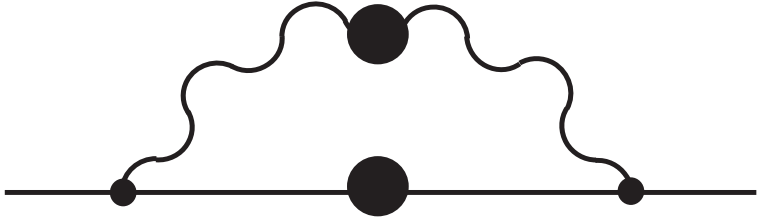

FIG. 5. Scalar self energy determining the damping rate of a scalar particle.

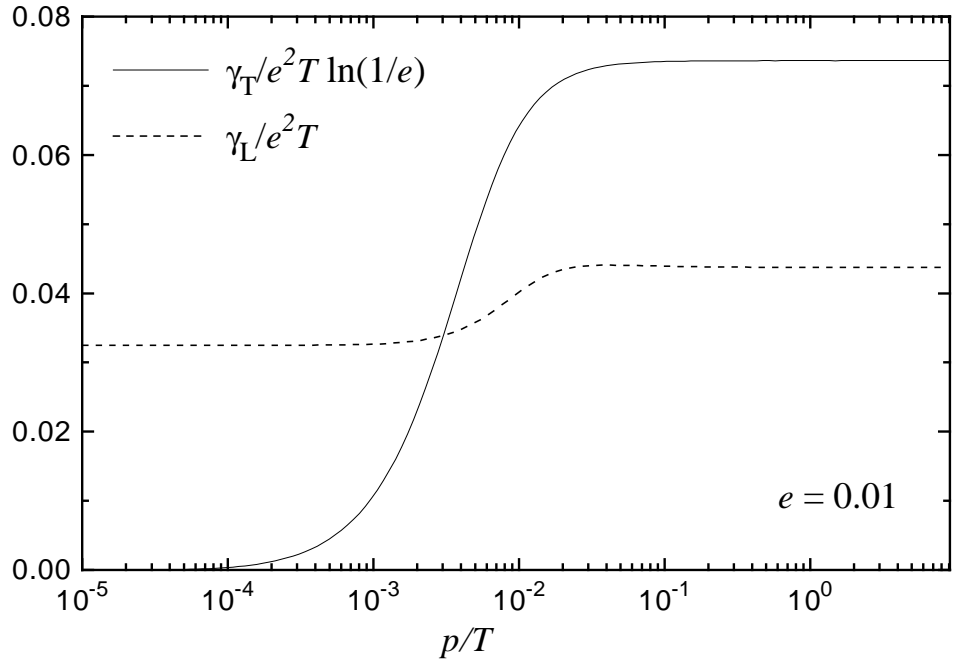

FIG. 6. Momentum dependence of the longitudinal and transverse parts of the scalar damping rate.

10